\documentclass[aps,prl,twocolumn,superscriptaddress,floatfix]{revtex4-2}
\usepackage{graphicx}
\usepackage{amsmath}
\usepackage{amssymb}
\usepackage{xcolor}
\usepackage[colorlinks=true,linkcolor=blue,citecolor=blue,urlcolor=blue]{hyperref}

\begin{document}

\title{Tunable Memory Effect in Dissipative Strongly Correlated Quantum Systems}
\author{Haowei Li}
\affiliation{Institute for Advanced Study, Tsinghua University, Beijing 100084, China}
\affiliation{Beijing Key Laboratory of Cold Atom Quantum Computation, Tsinghua University, Beijing 100084, China}
\author{Yu Chen}
\email{ychen@gscaep.ac.cn}
\affiliation{Graduate School of China Academy of Engineering Physics, Beijing 100193, China}
\author{Hui Zhai}
\email{hzhai@tsinghua.edu.cn}
\affiliation{Institute for Advanced Study, Tsinghua University, Beijing 100084, China}
\affiliation{Beijing Key Laboratory of Cold Atom Quantum Computation, Tsinghua University, Beijing 100084, China}
\affiliation{Hefei National Laboratory, Hefei 230088, China}
\date{\today}

\begin{abstract}

Strongly interacting quantum many-body systems subjected to non-Markovian dissipation pose a formidable challenge due to the interplay between strong correlation effects and memory effects. In this Letter, we develop a general theoretical framework to compute how a system observable responds to dissipation, which captures memory effects at short times and recovers the Markovian limit at longer times. Using this framework, we predict that, for a strongly correlated quantum critical state with critical exponent $\eta$, the short-time dynamics of a system observable always obeys a $t^{2\eta}$ scaling law. This emerges as a universal result from the interplay between strong correlation and memory effects, independent of the microscopic Hamiltonian of the system. We further reveal a crossover behavior of this scaling law to either $t^{2\eta-1}$ or linear-in-$t$ behavior beyond the memory time scale. We propose a concrete physical realization of a non-Markovian bath with tunable memory time using ultracold atoms, where our predictions can be straightforwardly verified in current experiments.

\end{abstract}

\maketitle

In closed systems, strong correlations pose a major challenge in quantum many-body physics because perturbation theory intrinsically fails, and strong interactions drive spectral weights to spread over a much broader high-energy regime, invalidating the quasi-particle picture~\cite{Sachdev2011,Giamarchi2004,VarmaNussinovVanSaarloos2002}. On the other hand, in open systems, a non-Markovian bath is generally more difficult to handle than a Markovian one because memory effects prevent the dissipation dynamics from being cast into a compact set of differential equations~\cite{BreuerPetruccione2002,DeVegaAlonso2017}. Therefore, studying strongly correlated quantum systems in a non-Markovian bath is a formidable challenge. So far, studies in this direction are largely limited to specific models, and general results are rare~\cite{Braggio2006,Cohen2011,Chen2017,Guo2018,YuChen2021,Su2021,Purkayastha2021,Flannigan2022,Fux2023,Chiriaco2023,Majumdar2023,Almeida2026}.

Quantum simulation platforms, such as ultracold atoms, are perfect for studying the interplay between strong correlation and non-Markovianity. The interaction effects between ultracold atoms are highly tunable, and the system can enter strongly correlated regimes by using Feshbach resonances~\cite{Chin2010} or by reducing dimensionality~\cite{Cazalilla2011,Guan2013}. Moreover, by optically controlling the atom-light interaction, we can control dissipation with high precision, not only its strength and duration, but also whether it is Markovian~\cite{Mueller2012,RivasHuelgaPlenio2014}. On the experimental side, it is now ready to integrate non-Markovianity with strong correlations to investigate the quantitative effects, calling for universal theoretical predictions for this interplay.

\begin{figure}[t]
\centering
\includegraphics[width=0.94\columnwidth]{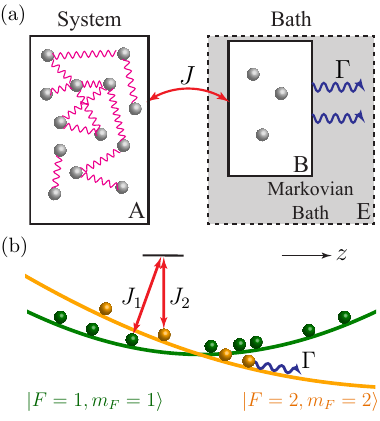} 
	\caption{(a) General setting. System $A$ with strongly interacting particles is coupled to a subsystem $B$, which is further coupled to a Markovian bath $E$. (b) Atoms in two different hyperfine spin states, denoted by green and yellow balls, are coupled by a Raman coupling. They experience a balanced or tilted harmonic trap, respectively.}
	\label{illustration}
\end{figure}
		
\textit{Physical Model.} We consider a general setting as shown in Fig.~\ref{illustration}(a). A system $A$ with strongly interacting particles is coupled to a subsystem $B$ via single-particle tunneling with strength $J$, where particles in $B$ are non-interacting and directly coupled to a Markovian bath $E$ with a dissipation strength $\Gamma$. This embedding of a finite-memory environment through auxiliary damped modes is closely related to the pseudomode construction of non-Markovian dynamics~\cite{Garraway1997,Pleasance2020}. The system can be generally described by a Hamiltonian with Langevin noise~\cite{GardinerZoller2004} as 
\begin{equation}
\hat{H}=\hat{H}_A(\{\hat{a}_i\})+J\sum_i(\hat{a}_i^\dag \hat{b}_i+\hat{b}_i^\dag \hat{a}_i)+\sqrt{2\Gamma}\sum_i(\hat{b}^\dag_i\hat{\xi}_i+\hat{\xi}^\dag_i\hat{b}_i)
\label{langevin}
\end{equation}
where $\hat{a}_i$ and $\hat{b}_i$ are the operators for modes $A$ and $B$, respectively. $\hat{\xi}_i$ is the Langevin noise operator of the Markovian bath that obeys the correlations 
\begin{align}
&\langle\hat{\xi}_i(t_1)\hat{\xi}^\dag_j(t_2)\rangle=\delta_{ij}\delta(t_1-t_2),\\
&\langle\hat{\xi}^\dag_i(t_1)\hat{\xi}^\dag_j(t_2)\rangle=\langle\hat{\xi}_i(t_1)\hat{\xi}_j(t_2)\rangle
=\langle\hat{\xi}^\dag_i(t_1)\hat{\xi}_j(t_2)\rangle=0.
\end{align}
In this work, we always focus on the regime $J\ll \Gamma$. 

{\it A Concrete Physical Realization.} As an example, we present a concrete physical realization of this model in Fig.~\ref{illustration}(b). Let us consider cold atoms in two different hyperfine spin states: one with $m_F=1$ and the other with $m_F=2$. For the $m_F=1$ state, the magnetic field gradient cancels the gravitational field along $\hat{z}$, so that atoms in this hyperfine state experience a balanced harmonic trap. For the $m_F=2$ state, these two fields do not cancel each other, resulting in a linear field gradient that drives atoms to leak out of the trap with a characteristic escape time $\tau=1/\Gamma$. Thus, $\Gamma$ can be tuned by adjusting the strength of the harmonic confinement relative to the gravitational field strength. These two states are coupled via a two-photon Raman process, which can be tuned by the laser intensity. Therefore, atoms in the two hyperfine states play the roles of system $A$ and $B$, respectively.

\textit{Summary of the Main Results.} Before proceeding to the detailed derivations, we summarize the main findings of this work. First, we identify two time scales, denoted by $t_0$ and $t_d$, respectively. $t_0 = 1/\Gamma$ is called the memory time and $t_d = \Gamma/J^2$ is called the dissipation time. The condition $J \ll \Gamma$ implies $t_0 \ll t_d$. In the limit $\Gamma, J \to \infty$ with $\Gamma/J^2$ fixed, the system under consideration recovers a fully Markovian environment. Consequently, we can consider two distinct time intervals, as far as the short dissipative dynamics is concerned: $t \ll t_0$ and $t_0 \ll t \ll t_d$. The first interval captures memory effects, while the second interval recovers the Markovian regime. 

\begin{table}[t]
	\caption{Scaling laws of the dissipation dynamics for the observable change $\delta\mathcal W(t)$ in different systems and time regimes.}
	\label{results}
	\begin{ruledtabular}
		\small
		\begin{tabular}{ccc}
			$\delta\mathcal W(t)$ & $t \ll t_0$ & $t_0 \ll t \ll t_d$ \\ 
			\hline
			Weakly correlated system, $[\hat{H}_A,\hat{W}]=0$ & $\propto t^{2}$         & $\propto t$ \\
			Weakly correlated system, $[\hat{H}_A,\hat{W}]\neq 0$ & $\propto t^{2}$         & $\propto t$ \\
			Critical system,  $[\hat{H}_A,\hat{W}]=0$   & $\propto t^{2\eta}$    & $\propto t$ \\
			Critical system, $[\hat{H}_A,\hat{W}]\neq 0$     & $\propto t^{2\eta}$    & $\propto t^{2\eta-1}$ \\
		\end{tabular}
	\end{ruledtabular}
\end{table}

We consider two different types of Hermitian observables $\hat W$ in system $A$, where $\hat W$ either commutes or does not commute with $\hat H_A$. When system $A$ is weakly correlated and possesses a well-defined quasi-particle description, its two-point correlation function behaves like a broadened delta function with finite width. In contrast, for strongly correlated systems, strong interaction effects push substantial spectral weight into high-energy tails, leading to a $1/\omega^{2\eta-1}$ critical behavior at large $\omega$~\cite{Giamarchi2004,ZinnJustin2007,Cazalilla2011}. For these two different cases, we respectively analyze how the change of the observable, $\delta\mathcal W(t)={\rm Tr}_A[\rho_A(t)\hat W]-{\rm Tr}_A[\rho_A(0)\hat W]$, scales with time $t$ after dissipation is turned on. We summarize the main results of this work, focusing on the non-Markovian regime $t \ll t_0$ and compare them with known results in the Markovian regime $t_0 \ll t \ll t_d$ in Table~\ref{results}.

We note that in the Markovian regime $t_0 \ll t$, if $[\hat{W},\hat{H}_A] = 0$, $\delta\mathcal W(t)$ always scales linearly with $t$, regardless of whether the system possesses a quasi-particle description~\cite{Supp}. If $[\hat{W},\hat{H}_A] \neq 0$ and the system is a critical state, $\delta\mathcal W(t)$ scales as $t^{2\eta-1}$, as predicted by non-Hermitian linear response theory~\cite{Pan2020} and confirmed experimentally~\cite{Zhao2025}.

Compared to the Markovian regime, the scaling laws of dissipation dynamics in the non-Markovian regime $t \ll t_0$ exhibit two nontrivial features: (1) Even when $[\hat{W},\hat{H}_A] = 0$, memory effects can convert linear $t$ behavior into $t^2$ or $t^{2\eta}$ behavior for weakly correlated or critical systems, respectively. (2) If $[\hat{W},\hat{H}_A] \neq 0$ and the system is a critical state, $\delta\mathcal W(t)$ scales as $t^{2\eta}$ instead of $t^{2\eta-1}$.

\textit{Theoretical Framework.} Our theory is based on a perturbative expansion to the second order of $J$. We start with the factorized initial state $\rho(0)=\rho_A^{(0)}\otimes\rho_B^{\rm vac}$, where $[\rho_A^{(0)},\hat H_A]=0$ and $\rho_B^{\rm vac}$ is the vacuum steady state of subsystem $B$. It is straightforward to show that, to the second order of $J$, the change of the observable can be derived as~\cite{Supp}
\begin{align}
	\delta\mathcal W(t)
	&=
	J^2 \sum_i
	\int_0^t\!\!dt_1
	\int_0^t\!\!dt_2\,
	e^{-\Gamma |t_1-t_2|} \notag\\
	&\Bigl\langle
	\hat a_i^\dagger(t_1)\hat W(t)\hat a_i(t_2)
	-\Theta(t_{1}-t_2)
	\hat W(t)\hat a_i^\dagger(t_1)\hat a_i(t_2)
	\notag\\
	&\quad
	-\Theta(t_{2}-t_1)
	\hat a_i^\dagger(t_1)\hat a_i(t_2)\hat W(t)
	\Bigr\rangle ,
	\label{eq:deltaW_general}
\end{align}
where $\hat W(t)$ and $\hat a_i(t)$ are interaction picture operators whose time dependence is generated by $\hat H_A$, and the expectation value is taken with respect to $\rho_A^{(0)}$. Here $\Theta(x)$ is the Heaviside step function, with $\Theta(x)=1$ for $x>0$, $\Theta(x)=0$ for $x<0$, and $\Theta(0)=1/2$. Eq.~\eqref{eq:deltaW_general} serves as a linear response theory to dissipation regardless of whether the environment is Markovian or not. The exponential kernel $e^{-\Gamma|t_1-t_2|}$ encodes non-Markovian dissipative memory with a time scale $t_0=1/\Gamma$. When $\Gamma\rightarrow\infty$, $(\Gamma/2)e^{-\Gamma |t|}$ approaches $\delta(t)$. Therefore, when $t\gg t_0$, Eq.~\eqref{eq:deltaW_general} recovers the Markovian limit
\begin{align} 
	\delta\mathcal W(t)\!=\!\frac{J^2}{\Gamma} \sum_i\!\int_0^t \!\!dt_1 \! \left\langle 2\hat a_i^\dagger(t_1)\hat W(t) \hat a_i(t_1)\!-\!\{\hat n_i(t_1),\!\hat W(t)\}\right\rangle,
	\label{eq:deltaWMK} 
\end{align}
where $\hat n_i(t_1)=\hat a_i^\dagger(t_1)\hat a_i(t_1)$. Eq.~\eqref{eq:deltaWMK} is precisely the non-Hermitian linear response theory derived earlier for a Markovian environment~\cite{Pan2020}, with the dissipative strength $J^2/\Gamma$, which defines a dissipation time scale $t_d=\Gamma/J^2$.

For later convenience, we introduce the Fourier transformation of Eq.~\eqref{eq:deltaW_general} as
\begin{equation}
	\delta\mathcal W(t)
	=
	J^2
	\int\frac{d\omega_1}{2\pi}
	\int\frac{d\omega_2}{2\pi}
	S^{(2)}_{W}(\omega_1,\omega_2)
	\mathcal K(\omega_1,\omega_2,t),
	\label{eq:general_spectral_response}
\end{equation}
where we define the two-frequency spectral density as
\begin{align}
		S^{(2)}_{W}(\omega_1,\omega_2)
		&=
		\sum_i
		\int_{-\infty}^{\infty}\!\!du
		\int_{-\infty}^{\infty}\!\!dv\,
		e^{-i(\omega_1u-\omega_2v)} \nonumber\\
		& \bigl[
		\langle
		\hat a_i^\dagger(u)\hat W\hat a_i(v)
		\rangle
		-
		\Theta(u-v)
		\langle
		\hat W\hat a_i^\dagger(u)\hat a_i(v)
		\rangle	\nonumber\\
	& -	\Theta(v-u)
		\langle
		\hat a_i^\dagger(u)\hat a_i(v)\hat W
		\rangle
		\bigr],
		\label{eq:SW2_def}
\end{align}
and the finite-time non-Markovian filter kernel as
\begin{equation}
	\mathcal K(\omega_1,\omega_2,t)
	=
	\int_0^t du
	\int_0^t dv\,
	e^{-\Gamma|u-v|}
	e^{-i\omega_1u}
	e^{i\omega_2v}.
	\label{eq:memory_filter_time}
\end{equation}

\textit{Case I: Conserved Observables.} For a conserved observable satisfying $[\hat H_A,\hat W]=0$ and $[\rho_A^{(0)},\hat W]=0$, Eq.~\eqref{eq:SW2_def} reduces to $S_{W}^{(2)}(\omega_1,\omega_2)
=
2\pi\delta(\omega_1-\omega_2)
S_{W}^{(1)}(\omega_1)$, where the one-frequency spectral density is given by
\begin{equation}
	S_W^{(1)}(\omega)
	=\sum_i\int_{-\infty}^{\infty}d\tau\,
	e^{-i\omega\tau}
	\left\langle[\hat a_i^\dagger(\tau),\hat W]\hat a_i(0)\right\rangle .
	\label{eq:SW1_def}
\end{equation}
Substituting this form into Eq.~\eqref{eq:general_spectral_response} gives
\begin{equation}
	\delta\mathcal W(t)
	=
	J^2
	\int\frac{d\omega}{2\pi}
	S_{W}^{(1)}(\omega)
	\mathcal K(\omega,\omega,t).
	\label{eq:conserved_response_from_SW2}
\end{equation}
It can be shown that $\mathcal K(\omega,\omega,t)$ behaves differently in the regime $t\ll t_0$ and $t_0\ll t\ll t_d$ as 
\begin{equation}
	\mathcal K(\omega,\omega,t)
	\simeq
	\begin{cases}
		\dfrac{2[1-\cos(\omega t)]}{\omega^2},
		& t\ll t_0, \\[1.2em]
		\dfrac{2\Gamma t}{\Gamma^2+\omega^2},
		& t_0\ll t\ll t_d .
	\end{cases}
	\label{eq:K_diag_asymptotic}
\end{equation}

\paragraph{Case IA: Conserved Observable in a Weakly Correlated System.} For weakly correlated systems, the one-frequency spectral density can be approximated by a Lorentzian peak as
\begin{equation}
	S_{W}^{(1)}(\omega)
	\approx
	\frac{2c\delta}
	{(\omega-\omega_0)^2+\delta^2},
	\label{eq:SW1_qp}
\end{equation}
where $c$ is the normalization factor, $\omega_0$ is the mode frequency, and $\delta$ is the quasi-particle linewidth.
Substituting Eqs.~\eqref{eq:K_diag_asymptotic} and \eqref{eq:SW1_qp} into Eq.~\eqref{eq:conserved_response_from_SW2}, one obtains
\begin{equation}
	\delta\mathcal W(t)
	\simeq
	\begin{cases}
		J^2 c t^2,
		& t\ll t_0, \\[1.0em]
		J^2 
		\dfrac{2c(\Gamma+\delta)}
		{\omega_0^2+(\Gamma+\delta)^2}t,
		& t_0\ll t\ll t_d .
	\end{cases}
	\label{eq:conserved_qp_result}
\end{equation}
Here, the $t^2$ behavior can be understood in terms of Rabi oscillations between systems $A$ and $B$ in the short time when tunneling dominates. 

\paragraph{Case IB: Conserved Observable in a Critical State.}
For quantum critical states, the spectral density $S_{W}^{(1)}(\omega)$ exhibits a universal power-law behavior~\cite{Giamarchi2004,ZinnJustin2007,Cazalilla2011} as
\begin{equation}
	S_{W}^{(1)}(\omega)
	\approx
	c(\omega-\omega_0)^{1-2\eta}\Theta\!\left(\omega-\omega_0\right),
	\label{eq:SW1_critical}
\end{equation}
where $1/2<\eta<1$ is the critical exponent. For $|\omega_0|\lesssim t_0^{-1}$, substituting Eqs.~\eqref{eq:K_diag_asymptotic} and \eqref{eq:SW1_critical} into Eq.~\eqref{eq:conserved_response_from_SW2} gives
\begin{equation}
	\delta\mathcal W(t)
	\simeq
	\begin{cases}
		\dfrac{J^2c}
		{2\mathcal G(2\eta+1)\sin(\pi\eta)}t^{2\eta},
		&  t\ll t_0,
		\\[1.2em]
		\dfrac{
			J^2c\,
			{\rm Im}\!\left[
			(\omega_0-i\Gamma)^{1-2\eta}
			\right]
		}{
			\sin\!\left[\pi(2\eta-1)\right]
		}
		t,
		&
		t_0\ll t\ll t_d,
	\end{cases}
	\label{eq:conserved_critical_result}
\end{equation}
where $\mathcal G(x)$ denotes the Euler gamma function.
\paragraph{Case II: Nonconserved Observables.} For a nonconserved observable, where $[\hat H_A,\hat W]\neq0$, the spectral density in Eq.~\eqref{eq:SW2_def} cannot be reduced to a one-frequency form. The asymptotic behavior of the two-frequency kernel in Eq.~\eqref{eq:memory_filter_time} is given by
\begin{equation}
	\mathcal K(\omega_1,\omega_2,t)
	\simeq
	\begin{cases}
		\dfrac{
			\left(e^{-i\omega_1 t}-1\right)
			\left(e^{i\omega_2 t}-1\right)
		}
		{\omega_1\omega_2},
		& t\ll t_0,
		\\[1.2em]
		\dfrac{2\Gamma}{\Gamma^2+\bar\omega^2}
		\dfrac{1-e^{-i(\omega_1-\omega_2)t}}{i(\omega_1-\omega_2)},
		& t_0\ll t\ll t_d 
	\end{cases}
	\label{eq:K_offdiag_asymptotic}
\end{equation}
where $\bar{\omega}=(\omega_1+\omega_2)/2$.

\paragraph{Case IIA: Nonconserved Observable in a Weakly Correlated System.}
For a weakly correlated system, we employ the Gaussian approximation, under which Wick's theorem factorizes the relevant higher-order correlation functions into products of two-point functions. Accordingly, the two-frequency spectral density can be approximated by a product of Lorentzian peaks as
\begin{equation}
	S_{W}^{(2)}(\omega_1,\omega_2)
	\approx
	\frac{
		4c\delta^2
	}{
		\left[(\omega_1-\omega_0)^2+\delta^2\right]
		\left[(\omega_2-\omega_0)^2+\delta^{2}\right]
	}.
	\label{eq:SW2_qp}
\end{equation}
Substituting Eqs.~\eqref{eq:K_offdiag_asymptotic} and \eqref{eq:SW2_qp} into
Eq.~\eqref{eq:general_spectral_response}, the leading response of a nonconserved observable is
\begin{equation}
	\delta\mathcal W(t)
	\simeq
	\begin{cases}
		J^2 c t^2,
		&
		t\ll t_0,
		\\[1.2em]
		J^2
		\dfrac{2c\Gamma}{\Gamma^2+\omega_0^2}\dfrac{1-e^{-2\delta t}}{2\delta},
		&
		t_0\ll t\ll t_d .
	\end{cases}
	\label{eq:nonconserved_qp_result}
\end{equation}
In the narrow-linewidth limit $\delta\lesssim t_d^{-1}$, one has $(1-e^{-2\delta t})/(2\delta)\simeq t$, giving a linear growth in the regime $t_0\ll t\ll t_d$.

\paragraph{Case IIB: Nonconserved Observable in a Critical State.}
For a critical state, we assume that the singular part of the two-frequency spectral density takes the factorized form following Wick contraction and threshold singularities of critical spectral functions~\cite{Giamarchi2004,Supp}
\begin{equation}
	S_{W}^{(2)}(\omega_1,\omega_2)
	\approx
	c
	\prod_{i=1}^{2}
	\left[
	(\omega_i-\omega_0)^{-\eta}
	\Theta\!\left(\omega_i-\omega_0\right)
	\right].
	\label{eq:SW2_critical}
\end{equation}
For $|\omega_0|\lesssim t_0^{-1}$, Eq.~\eqref{eq:general_spectral_response} can be evaluated using the two-frequency kernel in Eq.~\eqref{eq:K_offdiag_asymptotic}, giving
\begin{equation}
	\delta\mathcal W(t)
	\simeq
	\begin{cases}
		\displaystyle
		\dfrac{
			J^2 c
			[\mathcal G(1-\eta)]^2
		}{
			4\pi^2\eta^2
		}
		t^{2\eta},
		&
		t\ll t_0,
		\\[1.2em]
		\displaystyle
		\dfrac{
			J^2 c\Gamma
			[\mathcal G(1-\eta)]^2
		}{
			2\pi^2(2\eta-1)(\Gamma^2+\omega_0^2)
		}
		t^{2\eta-1},
		&
		t_0\ll t\ll t_d .
	\end{cases}
	\label{eq:nonconserved_critical_result}
\end{equation}

These results give Table~\ref{results}. We next illustrate them in two examples.
	
\begin{figure}[t]
		\includegraphics[width=\columnwidth]{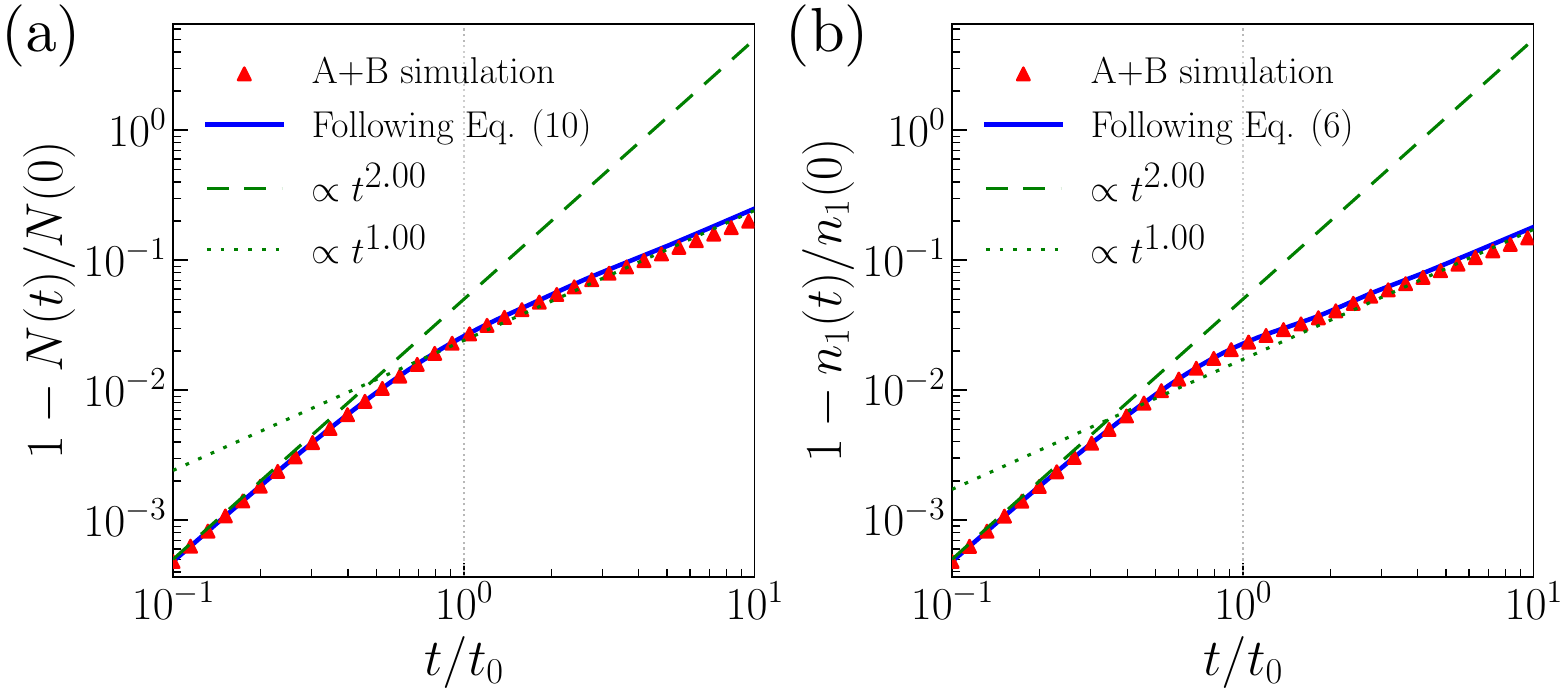}
		\caption{Dissipation dynamics in a two-mode Bose system for (a) the total population $N$ and (b) the mode population $n_1$. Red triangles show full master-equation simulations for $A+B$, and blue solid lines show the response theory results from Eq.~\eqref{eq:conserved_response_from_SW2} in (a) and Eq.~\eqref{eq:general_spectral_response} in (b). Green dashed (dotted) lines denote the short-time (long-time) asymptotes from Eq.~\eqref{eq:conserved_qp_result} in (a) and Eq.~\eqref{eq:nonconserved_qp_result} in (b). Gray vertical dotted lines mark $t=t_0$. Here $N_0=6$, $t_d/t_0=20$, $h_xt_0=2$, and $h_zt_0=1$.}
	\label{quasiparticle}
\end{figure}

\paragraph{Example I: Two-Mode Bose System.}
For weakly correlated systems, we consider a noninteracting two-mode Bose gas whose Hamiltonian is written as
	\begin{equation}
		\hat H_A
		=
		h_x
		\left(
		\hat a_1^\dagger \hat a_2
		+
		\hat a_2^\dagger \hat a_1
		\right)
		+
		2h_z\hat n_1.
	\end{equation} 
We initialize system $A$ in the infinite-temperature mixed state within the fixed-$N_0$ sector, for which $\langle \hat n_1\rangle=\langle \hat n_2\rangle=N_0/2$. We benchmark the second-order response theory against full master-equation simulations of the combined $A+B$ system, with $B$ coupled to the Markovian bath~\cite{Supp}. Fig.~\ref{quasiparticle} compares the full calculation with the prediction of Eq.~\eqref{eq:conserved_response_from_SW2} for the conserved total population $\hat N=\hat n_1+\hat n_2$ and that of Eq.~\eqref{eq:general_spectral_response} for the nonconserved mode population $\hat n_1$. The two approaches agree well for both observables, exhibiting a crossover from the short-time $t^2$ behavior to the long-time linear-in-$t$ behavior separated by the memory time scale $t_0$.

\begin{figure}[t] 
		\includegraphics[width=\columnwidth]{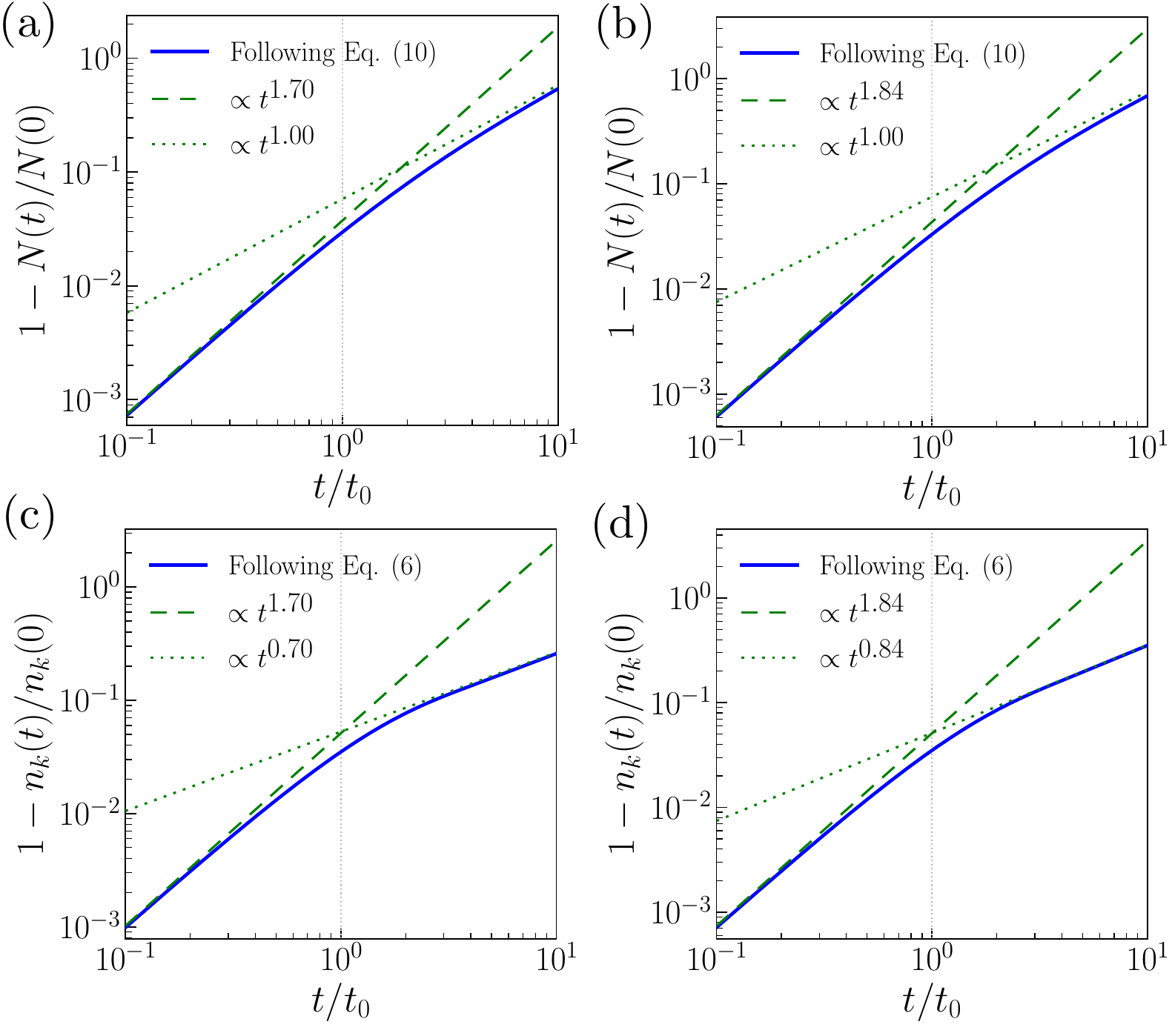} 
		\caption{Dissipation dynamics in a one-dimensional interacting Bose gas. (a), (b) Total population $N$; (c), (d) momentum occupation $n_k$ at $v_s|k|t_0=1$. The left (right) column corresponds to $\eta=0.85$ ($0.92$). Blue solid lines show the response theory results: Eq.~\eqref{eq:conserved_response_from_SW2} in (a), (b); Eq.~\eqref{eq:general_spectral_response} in (c), (d). Green dashed (dotted) lines denote the short-time (long-time) asymptotes: Eq.~\eqref{eq:conserved_critical_result} in (a), (b); Eq.~\eqref{eq:nonconserved_critical_result} in (c), (d). Gray vertical dotted lines mark $t=t_0$. Here $t_d/t_0=20$ and $\rho_0v_st_0=10$.}
		\label{luttinger} 
	\end{figure}	
\paragraph{Example II: One-Dimensional Interacting Bose Gas.}

As an example of critical states, we consider a one-dimensional interacting Bose gas~\cite{LiebLiniger1963}. We initialize system $A$ in its zero-temperature ground state within the fixed-$N_0$ sector. At low energies, it is characterized by the sound velocity $v_s$, the mean atomic density $\rho_0$, and the Luttinger parameter $K$, which fixes the critical exponent through $\eta=1-\frac{1}{4K}$~\cite{Giamarchi2004,Cazalilla2011}.
	
This many-body problem is beyond the capability of full numerical simulation. However, the relevant low-energy spectral densities can be computed using Luttinger-liquid theory~\cite{Giamarchi2004,Supp}, allowing us to calculate the response theory predictions from Eq.~\eqref{eq:general_spectral_response} and Eq.~\eqref{eq:conserved_response_from_SW2}. The results are shown in Fig.~\ref{luttinger}. Here we consider the conserved total atom population $\hat N=\int dx\hat a^\dagger(x)\hat a(x)$ and the nonconserved momentum population $\hat n_k=\hat a_k^\dagger\hat a_k$.

We show results for two representative critical exponents, $\eta=0.85$ in Fig.~\ref{luttinger}(a) and (c), and $\eta=0.92$ in Fig.~\ref{luttinger}(b) and (d). For both values of $\eta$, the total-number dynamics in Fig.~\ref{luttinger}(a) and (b) exhibit a crossover around $t_0$ from the non-Markovian $t^{2\eta}$ behavior to the linear-in-$t$ behavior. The momentum-occupation dynamics in Fig.~\ref{luttinger}(c) and (d) exhibit a crossover around $t_0$ from the non-Markovian $t^{2\eta}$ scaling to the long-time $t^{2\eta-1}$ behavior, confirming the asymptotic results discussed above.
	
\textit{Conclusion.} In summary, we have developed a general linear response theory for dissipation that applies irrespective of whether the environment is Markovian or non-Markovian. This theory captures the crossover from short-time non-Markovian dynamics, characterized by memory effects, to long-time Markovian dynamics. In particular, for a strongly correlated quantum critical state with critical exponent $\eta$, we predict that the non-Markovian dynamics obey a $t^{2\eta}$ scaling law, which is distinct from the $t^{2\eta-1}$ scaling law predicted and experimentally confirmed previously for Markovian dynamics in systems of ultracold atoms. We propose an experimental realization in ultracold atomic systems with tunable memory time, where the theoretical predictions of this work can be directly verified in current experiments.

\textit{Acknowledgments.} We thank Wenlan Chen for helpful discussions. This work was supported by the National Key Research and Development Program of China under Grant Nos.~2023YFA1406702 (H.Z.) and 2022YFA1405302 (Y.C.), No.~2025ZD0300400 (H.L.); the National Natural Science Foundation of China under Grant Nos.~12488301 (H.Z.), U23A6004 (H.Z.), 12174358 (Y.C.), U2330401 (Y.C.), and 12547168 (H.L.); the Innovation Program for Quantum Science and Technology under Grant No.~2021ZD0302005 (H.Z.); and the China National Postdoctoral Program for Innovative Talents under Grant No.~BX2026034 (H.L.).

\end{document}


\title{Supplemental Material for ``Tunable Memory Effect in Dissipative Strongly Correlated Quantum Systems''}
	
	\author{Haowei Li}
	\affiliation{Institute for Advanced Study, Tsinghua University, Beijing 100084, China}
	\affiliation{Beijing Key Laboratory of Cold Atom Quantum Computation, Tsinghua University, Beijing 100084, China}
	\author{Yu Chen}
	\email{ychen@gscaep.ac.cn}
	\affiliation{Graduate School of China Academy of Engineering Physics, Beijing 100193, China}
	\author{Hui Zhai}
	\email{hzhai@tsinghua.edu.cn}
	\affiliation{Institute for Advanced Study, Tsinghua University, Beijing 100084, China}
	\affiliation{Beijing Key Laboratory of Cold Atom Quantum Computation, Tsinghua University, Beijing 100084, China}
	\affiliation{Hefei National Laboratory, Hefei 230088, China}
	
	\maketitle
	
This Supplemental Material provides the derivations and model-specific details supporting the results presented in the main text. We first derive the response formula with a finite memory time, establish the linear-in-time scaling of conserved observables in the Markovian regime, and motivate the factorized two-frequency spectral density. We then analyze the two examples in the main text: for the two-mode Bose system, we present the response calculation and full-system benchmark; for the one-dimensional interacting Bose gas, we derive the Luttinger-liquid spectral density entering the total-number and momentum-occupation responses.

	\section{Perturbative derivation of observable dynamics with a finite memory time}
	\label{sec:supp_dissipative_response}
	
	We consider a system $A$ weakly coupled to auxiliary modes $B$ through the coherent Hamiltonian
	\begin{equation}
		\hat H_{AB}=\hat H_A(\{\hat a_i\})+J\sum_i(\hat a_i^\dagger\hat b_i+\hat b_i^\dagger\hat a_i).
	\end{equation}
	Each auxiliary mode is damped by a zero-temperature Markovian reservoir at rate $\Gamma$. We take the stationary factorized initial state
	$\rho(0)=\rho_A^{(0)}\otimes\rho_B^{\rm vac}$, where
	$[\rho_A^{(0)},\hat H_A]=0$ and $\rho_B^{\rm vac}$ is the auxiliary vacuum steady state. The auxiliary two-point correlations are
	\begin{equation}
		{\rm Tr}[\hat b_i(t_1)\hat b_j^\dagger(t_2)\rho_B^{\rm vac}]
		=\delta_{ij}e^{-\Gamma|t_1-t_2|},\qquad
		{\rm Tr}[\hat b_i^\dagger(t_1)\hat b_j(t_2)\rho_B^{\rm vac}]=0.
	\end{equation}
	In the interaction picture generated by $\hat H_A$ and the uncoupled damped auxiliary dynamics,
	\begin{equation}
		\hat H'_I(t)=J\sum_i[\hat a_i^\dagger(t)\hat b_i(t)+\hat b_i^\dagger(t)\hat a_i(t)].
	\end{equation}
	Expanding the evolution operator to second order,
	\begin{equation}
		\mathcal U_I(t)=1-i\int_0^t dt_1\,\hat H'_I(t_1)
		-\int_0^t dt_1\int_0^{t_1}dt_2\,\hat H'_I(t_1)\hat H'_I(t_2)+\mathcal O(J^3),
	\end{equation}
	Omitting terms beyond second order in $J$ and tracing out the auxiliary sector in $\rho_A^I(t)={\rm Tr}_B[\mathcal U_I(t)\rho(0)\mathcal U_I^\dagger(t)]$ gives $\delta\rho_A^I(t)\equiv\rho_A^I(t)-\rho_A^{(0)}$ as
	\begin{align}
		\delta\rho_A^I(t)=J^2\sum_i\int_0^t dt_1\int_0^t dt_2\,e^{-\Gamma|t_1-t_2|}
		\Big[&
		\hat a_i(t_2)\rho_A^{(0)}\hat a_i^\dagger(t_1)
		-\Theta(t_1-t_2)\hat a_i^\dagger(t_1)\hat a_i(t_2)\rho_A^{(0)}-\Theta(t_2-t_1)\rho_A^{(0)}\hat a_i^\dagger(t_1)\hat a_i(t_2)\Big],
	\end{align}
	where $\Theta(x)$ is the Heaviside step function, with $\Theta(x)=1$ for $x>0$, $\Theta(x)=0$ for $x<0$, and $\Theta(0)=1/2$.
	
	For a Hermitian system observable $\hat W$, we consider its dissipation-induced change $\delta\mathcal W(t)\equiv\mathcal{\rm Tr}_A[\rho_A(t)\hat W]-{\rm Tr}_A[\rho_A(0)\hat W]$. In the interaction picture, this change is equivalently $\delta\mathcal W(t)={\rm Tr}_A[\hat W(t)\delta\rho_A^I(t)]$. Hence
	\begin{align}
		\delta\mathcal W(t)
		=J^2\sum_i\int_0^t dt_1\int_0^t dt_2\,e^{-\Gamma|t_1-t_2|}
		\Big\langle&
		\hat a_i^\dagger(t_1)\hat W(t)\hat a_i(t_2)
		-\Theta(t_1-t_2)\hat W(t)\hat a_i^\dagger(t_1)\hat a_i(t_2)-\Theta(t_2-t_1)\hat a_i^\dagger(t_1)\hat a_i(t_2)\hat W(t)
		\Big\rangle .
		\label{eq:supp_deltaW_general}
	\end{align}
	All expectation values are evaluated in $\rho_A^{(0)}$, and the operators $\hat a_i(t)$ evolve under $\hat H_A$. Eq.~\eqref{eq:supp_deltaW_general} reproduces Eq.~(4) of the main text.

	\section{Linear scaling of conserved observables in the Markovian regime}
	\label{sec:supp_markov_conserved}

	Starting directly from Eq.~(4) of the main text, equivalently Eq.~\eqref{eq:supp_deltaW_general}, we take the Markovian limit $(\Gamma/2)e^{-\Gamma|t_1-t_2|}\to\delta(t_1-t_2)$. The response then becomes
	\begin{equation}
		\delta\mathcal W(t)
		=\frac{J^2}{\Gamma}\sum_i\int_0^t dt_1\,
		\left\langle
		2\hat a_i^\dagger(t_1)\hat W(t)\hat a_i(t_1)
		-\{\hat a_i^\dagger(t_1)\hat a_i(t_1),\hat W(t)\}
		\right\rangle .
		\label{eq:supp_markov_general}
	\end{equation}
	For a conserved observable $\hat W$ satisfying $[\hat H_A,\hat W]=0$. Let $U=e^{-i\hat H_A t_1}$, one has  $\hat W(t_1)=U^\dagger\hat W^\dagger U=\hat W$. Together with the stationarity condition $[\rho_A^{(0)},\hat H_A]=0$, this gives
	\begin{align}
		\left\langle
		2\hat a_i^\dagger(t_1)\hat W\hat a_i(t_1)
		-\{\hat a_i^\dagger(t_1)\hat a_i(t_1),\hat W\}
		\right\rangle
		={}&
		\operatorname{Tr}_A\Big[
		\rho_A^{(0)}
		\Big(
		2U^\dagger\hat a_i^\dagger U
		\hat W
		U^\dagger\hat a_iU
		-U^\dagger\hat a_i^\dagger\hat a_iU\hat W
		-\hat WU^\dagger\hat a_i^\dagger\hat a_iU
		\Big)
		\Big]
		\nonumber\\
		={}&
		\operatorname{Tr}_A\Big[
		\rho_A^{(0)}U^\dagger
		\Big(
		2\hat a_i^\dagger\hat W\hat a_i
		-\hat a_i^\dagger\hat a_i\hat W
		-\hat W\hat a_i^\dagger\hat a_i
		\Big)
		U
		\Big]
		\nonumber\\
		={}&
		\operatorname{Tr}_A\Big[
		U\rho_A^{(0)}U^\dagger
		\Big(
		2\hat a_i^\dagger\hat W\hat a_i
		-\hat a_i^\dagger\hat a_i\hat W
		-\hat W\hat a_i^\dagger\hat a_i
		\Big)
		\Big]
		\nonumber\\
		={}&
		\left\langle
		2\hat a_i^\dagger\hat W\hat a_i
		-\{\hat a_i^\dagger\hat a_i,\hat W\}
		\right\rangle .
		\label{eq:supp_stationary_integrand}
	\end{align}
	Therefore, the integrand in
	Eq.~\eqref{eq:supp_markov_general} is independent of $t_1$, and direct
	integration gives
	\begin{equation}
		\delta\mathcal W(t)
		=\frac{J^2t}{\Gamma}\sum_i
		\left\langle
		2\hat a_i^\dagger\hat W\hat a_i
		-\{\hat a_i^\dagger\hat a_i,\hat W\}
		\right\rangle,
		\label{eq:supp_markov_conserved}
	\end{equation}
	which indicates the response of an observable
	conserved by $\hat H_A$ is linear in $t$ for $t_0\ll t\ll t_d$.
	
	\section{Motivation for the factorized two-frequency spectral density}
	\label{sec:supp_spectral_response}
	
	Here we give a simple motivation for the factorized form used in Case IIB. Consider a one-body observable
	\begin{equation}
		\hat W(t)=\sum_{\alpha,\beta}w_{\alpha\beta}
		\hat a_\alpha^\dagger(t)\hat a_\beta(t),
		\label{eq:supp_general_bilinear_W}
	\end{equation}
	where the indices may denote internal states, position, or momentum. In the definition of $S_W^{(2)}(\omega_1,\omega_2)$ in the main text, $u$ and $v$ are the two time variables integrated in the double Fourier transform and are conjugate to $\omega_1$ and $\omega_2$, respectively. Because the initial state is stationary under $\hat H_A$, the time origin can be chosen at the observable, so that $\hat W=\hat W(0)$, while the two external operators are $\hat a_i^\dagger(u)$ and $\hat a_i(v)$. The definition therefore contains four-operator correlation functions. Within the pair-contraction approximation, directly contracting these two external operators leaves a factor proportional to $\langle\hat W\rangle$. The same factor appears in all three terms of $S_W^{(2)}$ and cancels because $1-\Theta(u-v)-\Theta(v-u)=0$.
	
	In the remaining cross-contractions, $\hat a_i^\dagger(u)$ is paired with one field in $\hat W(0)$, while $\hat a_i(v)$ is paired with the other. A representative term is
	\begin{equation}
		\langle\hat a_i^\dagger(u)\hat a_\beta(0)\rangle
		\langle\hat a_\alpha^\dagger(0)\hat a_i(v)\rangle .
		\label{eq:supp_cross_pairing}
	\end{equation}
	Before the double Fourier transform, the $u$ and $v$ dependences therefore reside in separate two-point functions; after the transform, they generate the separate $\omega_1$ and $\omega_2$ dependences of $S_W^{(2)}(\omega_1,\omega_2)$. If the Fourier transform of each two-point function has the same dominant threshold behavior
	\begin{equation}
		g_{\rm edge}(\omega)
		=(\omega-\omega_0)^{-\eta}\Theta(\omega-\omega_0).
	\end{equation}
	then the leading singularity of the two-frequency spectral density is the product of the two edges,
	\begin{equation}
		S_{W,{\rm sing}}^{(2)}(\omega_1,\omega_2)\approx
		c\prod_{l=1}^{2}[(\omega_l-\omega_0)^{-\eta}\Theta(\omega_l-\omega_0)].
		\label{eq:supp_SW2_factorized_sing}
	\end{equation}
	Here $c$ collects the smooth prefactors and ordering-dependent weights. The same reasoning extends to observables containing four or more field operators: within the pair-contraction approximation, the additional fields only modify $c$, while the $u$ and $v$ dependences remain in two separate two-point functions. Eq.~\eqref{eq:supp_SW2_factorized_sing} describes only the leading threshold singularity, rather than the full two-frequency spectrum. It applies when the remaining frequency dependence is smooth near the selected threshold. The momentum-occupation example below gives an explicit realization of this factorized structure.
	
	\section{Response of a Two-Mode Bose System}
	\label{sec:spinor_model}
	
	To provide details for Example I in the main text, we apply the spectral representation to the noninteracting two-mode Bose system, whose Hamiltonian is
	\begin{equation}
		\hat H_A=h_x(\hat a_1^\dagger\hat a_2+\hat a_2^\dagger\hat a_1)
		+2h_z\hat a_1^\dagger\hat a_1 .
		\label{eq:S_spinor_H}
	\end{equation}
	With $h=\sqrt{h_x^2+h_z^2}$, its eigenenergies are
	$\epsilon_\sigma=h_z+\sigma h$, and the Hamiltonian can be written as
	$\hat H_A=\epsilon_+\hat\alpha_+^\dagger\hat\alpha_+
	+\epsilon_-\hat\alpha_-^\dagger\hat\alpha_-$,
	where $\hat\alpha_\pm$ are two quasi-particles with energies $\epsilon_\pm$.
	We use the infinite-temperature state in the fixed-$N_0$ sector,
	\begin{equation}
		\rho_A^{(0)}
		=\frac{1}{N_0+1}
		\sum_{m=0}^{N_0}
		|m,N_0-m\rangle\langle m,N_0-m| .
		\label{eq:S_spinor_initial_state}
	\end{equation}
	It commutes with $\hat H_A$ and satisfies
	\begin{equation}
		\langle\hat\alpha_{\sigma'}^\dagger\hat\alpha_\sigma\rangle_0
		=\frac{N_0}{2}\delta_{\sigma\sigma'},
		\qquad
		\langle\hat n_1(t)\rangle_0
		=\langle\hat n_2(t)\rangle_0
		=\frac{N_0}{2}.
		\label{eq:S_spinor_initial_observables}
	\end{equation}
	The full $A+B$ calculation shown in the main text solves
	\begin{align}
		\frac{d\rho_{AB}}{dt}
		=-i[\hat H_{AB},\rho_{AB}]+\Gamma\sum_{i=1}^{2}
		\left(2\hat b_i\rho_{AB}\hat b_i^\dagger
		-\{\hat b_i^\dagger\hat b_i,\rho_{AB}\}\right),
		\label{eq:S_spinor_master_equation}
	\end{align}
	with $N_0=6$ and
	\begin{equation}
		\rho_{AB}(0)=\rho_A^{(0)}\otimes
		|0,0\rangle_B\langle0,0|.
		\label{eq:S_spinor_AB_initial_state}
	\end{equation}
	
	For the conserved total atom number
	$\hat N=\hat n_1+\hat n_2$, the one-frequency spectral density and response are
	\begin{equation}
		S_N^{(1)}(\omega)
		=-\pi N_0\sum_{\sigma=\pm1}\delta(\omega-\epsilon_\sigma),
		\qquad
		\delta N(t)
		=-\frac{J^2N_0}{2}
		\sum_{\sigma=\pm1}\mathcal K(\epsilon_\sigma,\epsilon_\sigma,t).
		\label{eq:S_spinor_SN}
	\end{equation}
	This is the zero-linewidth limit of the quasi-particle form in the main text.
	For $|\epsilon_\sigma|\sim\Gamma$, the diagonal kernel obeys
	$\mathcal K(\epsilon_\sigma,\epsilon_\sigma,t)\simeq t^2$ for $t\ll t_0$
	and $2\Gamma t/(\Gamma^2+\epsilon_\sigma^2)$ for
	$t_0\ll t\ll t_d$. Hence
	\begin{equation}
		\delta N(t)\simeq
		\begin{cases}
			-J^2N_0t^2,
			& t\ll t_0,\\[0.6em]
			-J^2N_0\Gamma t
			\displaystyle\sum_{\sigma=\pm1}
			\frac{1}{\Gamma^2+\epsilon_\sigma^2},
			& t_0\ll t\ll t_d.
		\end{cases}
		\label{eq:S_spinor_deltaN_asymptotic}
	\end{equation}
	The component population $\hat{n}_1$ is nonconserved for $h_x\neq0$.
	In the quasi-particle basis,
	\begin{align}
		\hat n_1(t)
		=\frac{1}{2}\sum_{\sigma=\pm1}
		\left(1+\sigma\frac{h_z}{h}\right)
		\hat\alpha_\sigma^\dagger\hat\alpha_\sigma+\frac{h_x}{2h}
		\left(
		e^{2iht}\hat\alpha_+^\dagger\hat\alpha_-
		+
		e^{-2iht}\hat\alpha_-^\dagger\hat\alpha_+
		\right).
		\label{eq:S_spinor_n1}
	\end{align}
	For the stationary state in Eq.~\eqref{eq:S_spinor_initial_state}
	and symmetric loss channels, the off-diagonal coherence does not contribute.
	The two-frequency spectral density therefore reduces to
	\begin{equation}
		S_{n_1}^{(2)}(\omega_1,\omega_2)
		=-\pi^2N_0\sum_{\sigma=\pm1}
		\left(1+\sigma\frac{h_z}{h}\right)
		\delta(\omega_1-\epsilon_\sigma)
		\delta(\omega_2-\epsilon_\sigma).
		\label{eq:S_spinor_Sn1}
	\end{equation}
	Substituting Eq.~\eqref{eq:S_spinor_Sn1} into the general two-frequency
	response formula gives
	\begin{equation}
		\delta n_1(t)
		=-\frac{J^2N_0}{4}\sum_{\sigma=\pm1}
		\left(1+\sigma\frac{h_z}{h}\right)
		\mathcal K(\epsilon_\sigma,\epsilon_\sigma,t).
		\label{eq:S_spinor_deltan1}
	\end{equation}
	Applying the same kernel limits as for the total atom number, we obtain
	\begin{equation}
		\delta n_1(t)\simeq
		\begin{cases}
			-\dfrac{J^2N_0}{2}t^2,
			& t\ll t_0,\\[0.8em]
			-\dfrac{J^2N_0\Gamma t}{2}
			\displaystyle\sum_{\sigma=\pm1}
			\left(1+\sigma\frac{h_z}{h}\right)
			\frac{1}{\Gamma^2+\epsilon_\sigma^2},
			& t_0\ll t\ll t_d.
		\end{cases}
		\label{eq:S_spinor_deltan1_asymptotic}
	\end{equation}
	Thus, the responses of both the conserved total atom number and the
	nonconserved component population exhibit a crossover from the short-time
	$t^2$ behavior to the long-time linear-in-$t$ behavior, separated by the
	memory time scale $t_0$, in agreement with the results shown in the main text.
	
	\section{Response of a One-Dimensional Interacting Bose Gas}
	\label{sec:ll_model}
	
	To provide details for Example II in the main text, we consider the same one-dimensional interacting Bose gas, described microscopically by the Lieb--Liniger Hamiltonian,
	\begin{equation}
		\hat H_A
		=
		\int dx
		\left[
		\frac{1}{2m}
		\partial_x\hat a^\dagger(x)
		\partial_x\hat a(x)
		+
		\frac{g}{2}
		\hat a^\dagger(x)\hat a^\dagger(x)
		\hat a(x)\hat a(x)
		\right].
		\label{eq:S_LL_H}
	\end{equation}
	Here $m$ is the atomic mass, and $g$ is the one-dimensional contact-interaction strength.
	For a system of length $L$ containing a fixed atom number $N_0$, the mean density is $\rho_0=N_0/L$. Using bosonization, the field operator admits the standard long-wavelength form
	\begin{equation}
		\hat a(x)
		\simeq\sqrt{\rho_0 - \frac{1}{\pi}\partial_x\phi(x)}\,
		e^{i\theta(x)},
		\label{eq:S_bosonization_field}
	\end{equation}
	where $\phi(x)$ and $\theta(x)$ are the density and phase fields, respectively. These fields satisfy $[\phi(x),\partial_y\theta(y)]=i\pi\delta(x-y)$.
	In terms of these fields, the low-energy Hamiltonian, parameterized by the sound velocity $v_s$ and Luttinger parameter $K$, is
	\begin{equation}
		\hat H_{\rm LL}
		=
		\frac{v_s}{2\pi}
		\int dx
		\left[
		K(\partial_x\theta)^2
		+
		\frac{1}{K}(\partial_x\phi)^2
		\right],
		\label{eq:S_LL_HLL}
	\end{equation}
	The parameters $v_s$ and $K$ are determined by the equation of state; in particular, expressing the ground-state energy as $E(L)$ at fixed atom number $N_0$, one may write
	\begin{equation}
		v_s = \sqrt{\frac{L}{m\rho_0}\frac{\partial^2 E}{\partial L^2}},\qquad
		K = \pi \rho_0 \sqrt{\frac{\rho_0}{mL\,\partial^2 E/\partial L^2}}.
	\end{equation}
	
	At zero temperature, the single-particle correlator takes the scaling form~\cite{Giamarchi2004,Cazalilla2011},
	\begin{equation}
		G^<(x,\tau)\equiv\langle\hat a^\dagger(x,\tau)\hat a(0,0)\rangle
		\simeq\rho_0\prod_{\sigma=\pm1}
		\left[1-i\frac{\tau}{t_{\rm UV}}+i\sigma\rho_0x\right]^{\eta-1}.
		\label{eq:S_LL_G_less}
	\end{equation}
	Here the critical exponent $\eta=1-1/(4K)$, and the ultraviolet cutoff time $t_{\rm UV}=(\rho_0v_s)^{-1}$.
	
	\subsection{Response of the conserved total atom number}
	
	For $\hat N=\int dx\,\hat a^\dagger(x)\hat a(x)$, the commutator $[\hat N,\hat a(x)]=-\hat a(x)$ reduces the one-frequency spectral density to
	\begin{equation}
		S_N^{(1)}(\omega)=-\int_{-\infty}^{\infty}d\tau\,e^{-i\omega\tau}
		\int dx\,\langle\hat a^\dagger(x,\tau)\hat a(x,0)\rangle .
		\label{eq:S_LL_SN_from_def_2}
	\end{equation}
	For a translationally invariant system, Eq.~\eqref{eq:S_LL_G_less} gives
	\begin{equation}
		\int dx\,\langle\hat a^\dagger(x,\tau)\hat a(x,0)\rangle
		\simeq N_0(1-i\tau/t_{\rm UV})^{2\eta-2}.
	\end{equation}
	A Lehmann representation confirms that this Fourier convention places positive particle-removal energies at $\omega>0$. Using
	\begin{equation}
		(1-i\tau/t_{\rm UV})^{-(2-2\eta)}
		=\frac{1}{\mathcal G(2-2\eta)}\int_0^\infty ds\,s^{1-2\eta}e^{-s+is\tau/t_{\rm UV}},
	\end{equation}
	we obtain
	\begin{equation}
		S_N^{(1)}(\omega)=
		-\frac{2\pi N_0t_{\rm UV}}{\mathcal G(2-2\eta)}
		(\omega t_{\rm UV})^{1-2\eta}e^{-\omega t_{\rm UV}}\Theta(\omega).
		\label{eq:S_LL_SN}
	\end{equation}
	Thus, for $|\omega|t_{\rm UV}\ll1$,
	\begin{equation}
		S_N^{(1)}(\omega)\simeq c_\eta\omega^{1-2\eta}\Theta(\omega),
		\qquad c_\eta=-\frac{2\pi N_0t_{\rm UV}^{2-2\eta}}{\mathcal G(2-2\eta)}.
		\label{eq:S_LL_SN_low}
	\end{equation}
	Here $\mathcal G$ is the Euler gamma function. Substitution into the conserved-observable response formula of the main text yields
	\begin{equation}
		\delta N(t)\simeq
		\begin{cases}
			\dfrac{J^2c_\eta}{2\mathcal G(2\eta+1)\sin(\pi\eta)}\,t^{2\eta},
			& t_{\rm UV}\ll t\ll t_0,\\[1.2em]
			\dfrac{J^2c_\eta\Gamma^{1-2\eta}}{2\sin(\pi\eta)}\,t,
			& t_0\ll t\ll t_d .
		\end{cases}
	\end{equation}
	The conserved response therefore crosses over from the non-Markovian $t^{2\eta}$ behavior to linear-in-$t$ behavior in the Markovian regime.
	
	\subsection{Response of the momentum occupation}
	
	We next consider the momentum occupation $\hat n_k(t)=\hat a_k^\dagger(t)\hat a_k(t)$, where $\hat a_k(t)=L^{-1/2}\int dy\,e^{iky}\hat a(y,t)$. This observable is nonconserved and is therefore governed by the two-frequency spectral density. Using translational invariance to carry out the center-of-mass average, direct substitution gives
	\begin{align}
		S_{n_k}^{(2)}(\omega_1,\omega_2)
		=&
		\int dx
		\int dy\,e^{-iky}
		\int_{-\infty}^{\infty}du
		\int_{-\infty}^{\infty}dv\,
		e^{-i\omega_1u+i\omega_2v}
		\Big[
		\left\langle
		\hat a^\dagger(x,u)\hat a^\dagger(y,0)\hat a(0,0)\hat a(x,v)
		\right\rangle\notag\\
		&
		-
		\Theta(u-v)
		\left\langle
		\hat a^\dagger(y,0)\hat a(0,0)\hat a^\dagger(x,u)\hat a(x,v)
		\right\rangle
		-
		\Theta(v-u)
		\left\langle
		\hat a^\dagger(x,u)\hat a(x,v)\hat a^\dagger(y,0)\hat a(0,0)
		\right\rangle
		\Big].
		\label{eq:S_LL_Snk_from_def}
	\end{align}
	The four-point functions are evaluated using Wick's theorem and can be reduced to
	\begin{align}
		S_{n_k}^{(2)}(\omega_1,\omega_2)
		&\simeq
		-
		\int_{-\infty}^{\infty}du
		\int_{-\infty}^{\infty}dv\,
		e^{-i\omega_1u+i\omega_2v}
		\Big[
		\Theta(u-v)\,
		G^<(k,-v)A(k,u)
		+
		\Theta(v-u)\,
		G^<(k,u)A(k,-v)
		\Big],
		\label{eq:S_LL_Snk_GA}
	\end{align}
	where $
	G^<(k,\tau)
	=
	\int dx\,e^{-ikx}
	\left\langle
	\hat a^\dagger(x,\tau)\hat a(0,0)
	\right\rangle$, 
	and	$A(k,\tau)=G^>(k,\tau)-G^<(k,\tau)$.
	In the low-energy window relevant to the response, we approximate the lesser Green function by the spectral function weighted by the initial momentum occupation, $G^<(k,\tau)\simeq n_k(0)A(k,\tau)$. With $A(k,\omega)=\int d\tau\,e^{-i\omega\tau}A(k,\tau)$, the two products in Eq.~\eqref{eq:S_LL_Snk_GA} coincide and $\Theta(v-u)+\Theta(u-v)=1$. The singular part therefore factorizes as
	\begin{equation}
		S_{n_k}^{(2)}(\omega_1,\omega_2)
		\simeq-n_k(0)A(k,\omega_1)A(k,\omega_2).
		\label{eq:S_nk_SW2_factorized}
	\end{equation}

Eq.~\eqref{eq:S_LL_G_less} fixes the normalization consistently with the threshold representation. With $q=\Delta_kt_{\rm UV}=|k|/\rho_0$ and $\Delta_k=v_s|k|$, its Fourier transform is
	\begin{equation}
		G^<(k,\omega)\equiv\int d\tau\,e^{-i\omega\tau}G^<(k,\tau)
		=\frac{2^{2\eta+1}\pi^2t_{\rm UV}^{1-2\eta}}
		{\mathcal G(1-\eta)^2}e^{-\omega t_{\rm UV}}
		(\omega^2-\Delta_k^2)^{-\eta}\Theta(\omega-\Delta_k).
		\label{eq:S_LL_G_less_kw}
	\end{equation}
	This result follows by expressing the two factors in Eq.~\eqref{eq:S_LL_G_less} in Laplace form. The $x$ and $\tau$ integrals then set the two positive Laplace variables to $(\omega t_{\rm UV}\mp q)/2$, yielding the stated threshold and prefactor. At equal time, the same equation gives
	$n_k(0)=2\sqrt{\pi}
		\left(\frac{q}{2}\right)^{1/2-\eta}K_{1/2-\eta}(q)/\mathcal G(1-\eta)$.
	Within the approximation $G^<(k,\tau)\simeq n_k(0)A(k,\tau)$ used above, Eq.~\eqref{eq:S_LL_G_less_kw} therefore fixes the spectral function on this branch as
	\begin{equation}
		A(k,\omega)\simeq
		\frac{2^{2\eta+1}\pi^2t_{\rm UV}^{1-2\eta}}
		{n_k(0)\mathcal G(1-\eta)^2}
		e^{-\omega t_{\rm UV}}(\omega^2-\Delta_k^2)^{-\eta}
		\Theta(\omega-\Delta_k).
		\label{eq:S_LL_Akomega}
	\end{equation}
	Thus the branch retained in Eq.~\eqref{eq:S_nk_SW2_factorized} is fixed by Eq.~\eqref{eq:S_LL_G_less}; no symmetric continuation in frequency is assumed. Evaluating the smooth exponential at the threshold gives
	\begin{equation}
		S_{n_k}^{(2)}(\omega_1,\omega_2)
		\simeq c_{k,\eta}\prod_{a=1}^{2}
		\left[(\omega_a^2-\Delta_k^2)^{-\eta}\Theta(\omega_a-\Delta_k)\right],
		\label{eq:S_nk_SW2}
	\end{equation}
	where
	\begin{align}
		c_{k,\eta}
		=-\frac{2^{4\eta+2}\pi^4t_{\rm UV}^{2-4\eta}e^{-2q}}
		{n_k(0)\mathcal G(1-\eta)^4}=-\frac{2\pi^3n_k(0)t_{\rm UV}^{2-2\eta}e^{-2q}(2\Delta_k)^{2\eta}}
		{qK_{\frac12-\eta}(q)^2\mathcal G(1-\eta)^2}.
		\label{eq:S_LL_cketa}
	\end{align}
	Here $K_\nu$ is the modified Bessel function of the second kind, and the second equality follows from the equal-time value of Eq.~\eqref{eq:S_LL_G_less} displayed above.
	Near the removal threshold, $\omega_a^2-\Delta_k^2\simeq2\Delta_k(\omega_a-\Delta_k)$, so the coefficient of the local critical form is $c_{k,\eta}/(2\Delta_k)^{2\eta}$. This edge has the positive-frequency convention used in the main text.

	Eqs.~\eqref{eq:S_nk_SW2_factorized}--\eqref{eq:S_LL_cketa} define a normalized Wick-factorization approximation. For an interacting Luttinger liquid, connected four-vertex terms can renormalize the amplitude or add other branches; the exponents below remain unchanged provided that they introduce no stronger threshold singularity. We also assume that the edge power law remains applicable over the low-energy frequencies selected by the response kernel. In the non-Markovian window this continues the local form to detunings of order $1/t$, even when $1/t\gtrsim\Delta_k$.

	For $\Delta_k\lesssim\Gamma$ and $t_{\rm UV}\ll t_0$, substituting Eqs.~\eqref{eq:S_nk_SW2} and \eqref{eq:S_LL_cketa} into the two-frequency response gives
	\begin{equation}
		\delta n_k(t)\simeq
		\begin{cases}
			-\dfrac{\pi J^2n_k(0)t_{\rm UV}^{2-2\eta}e^{-2q}}
			{2\eta^2qK_{\frac12-\eta}(q)^2}\,t^{2\eta},
			& t_{\rm UV}\ll t\ll t_0,\\[1.2em]
			-\dfrac{\pi J^2n_k(0)t_{\rm UV}^{2-2\eta}e^{-2q}\Gamma}
			{(2\eta-1)qK_{\frac12-\eta}(q)^2(\Gamma^2+\Delta_k^2)}\,t^{2\eta-1},
			& t_0\ll t\ll t_d .
		\end{cases}
		\label{eq:S_LL_deltank_asymptotic}
	\end{equation}
	Thus the nonconserved momentum occupation crosses over from the non-Markovian $t^{2\eta}$ behavior to the Markovian $t^{2\eta-1}$ behavior.